\documentclass[showpacs,twocolumn,superscriptaddress,amsmath,amssymb,aps,pre]{revtex4}
\usepackage{graphicx}
\usepackage{dcolumn}
\usepackage{bm}
\usepackage{color}
\usepackage{amsmath,amssymb}
\newcommand{\req}[1]{Eq.~(\ref{#1})}
\newcommand{\avg}[1]{\langle #1\rangle}

\newcommand{\fig}[1]{Fig.~\ref{#1}}

\begin{document}


\title{Self-organization in social tagging systems}

\author{Chuang Liu}
 \affiliation{School of Business, East China University of Science and Technology, Shanghai 200237, China} %
 \affiliation{Engineering Research Center of Process Systems Engineering (Ministry of Education), East China University of Science and Technology, Shanghai 200237, China} %
\author{Chi Ho Yeung}
\thanks{chbyeung@gmail.com}
 \affiliation{Department of Physics, University of Fribourg, Fribourg CH-1700, Switzerland}
\author{Zi-Ke Zhang}
\thanks{zhangzike@gmail.com}
 \affiliation{Department of Physics, University of Fribourg, Fribourg CH-1700, Switzerland}
  \affiliation{Web Sciences Center, University of Electronic Science and Technology of China - Chengdu 610054, PRC}

\date{\today}

\begin{abstract}
Individuals often imitate each other to fall into the typical group,
leading to a self-organized state of typical behaviors in a community.
In this paper, we model self-organization in social tagging systems and
illustrate the underlying interaction and dynamics.
Specifically, we introduce a model in which individuals adjust
their own tagging tendency to imitate the average tagging tendency.
We found that when users are of low
confidence,
they tend to imitate others and lead to a self-organized state with active tagging. On
the other hand, when
users are of high confidence and are stubborn for changes,
tagging becomes inactive. We observe a
phase transition at a critical level of user confidence when the
system changes from one regime
to the other. The distributions of post length obtained from the model are compared to
real data which show good agreements.
\end{abstract}


\maketitle


\section{Introduction}
Self-organization is an interesting phenomenon observed in various
areas including network growth \cite{Dorogovtsev2001},
traffic jams \cite{Chowdhury1999} and resource allocation
\cite{yeung2010}. 
In social systems, 
individuals often imitate each other
through interaction and observation, 
to become more typical in the community. 
Such dynamics results in a steady state in
which most individuals adopt the typical practice by learning from
each other. 
In online communities,
self-organization is further facilitated by
the recent advent of Web 2.0 social applications, which encourage
Internet users to interact with peers. By interacting with each
other, users self-organize and lead to a state of typical behaviors.

In resource sharing applications,
tags are practical to facilitate the
search and management of resources \cite{Golder2006,Cattuto2007}.
Tags are usually simple labels and annotations which help users to have preliminary understanding
of the content before collecting the resources.
Recently, \emph{tagging systems} are implemented in popular applications including 
\emph{delicious.com}, \emph{flickr.com}
and \emph{citeulike.org}.
To well organize their resources,
users assign tags with their bookmark,
pictures or Bibtex files.
By browsing through tags, users are able to find other users who
share similar interests. Tags thus reflect user behaviors and
preferences, and with which ones can easily search, collaborate and
form communities with others \cite{Sen2006}.

Tagging systems are studied extensively in recent years, but the
underlying interaction and dynamics among tag users are still
unclear. Mathematically, tagging systems are composed of fundamental
units of user-resource-{tag} triples
\cite{Cattuto2007,Ghoshal2009,Zlatic2009}, and each tagging action
constitutes one or several hyper-links in a tripartite graph. Such
user-resource-tag relations are often referred to as
\emph{folksonomy}. Examples include the use of keywords or PACS
numbers in academic papers,
which also helps to reveal the structure of citation networks \cite{Palla2008, ZhangZK2008}. 
However, how
similar papers influence each other on the choice of keywords is still
an open question. 
To reveal the tagging dynamics, 
Cattuto \emph{et al} \cite{Cattuto2009} suggested to consider the process of social
annotation as a collective yet uncoordinated exploration
of the underlying semantic space through a series of random walks. 
In Ref. \cite{Lambiotte2006}, Lambiotte \emph{et al} modeled folksonomy
in terms of tripartite graphs. Zhang and Liu \cite{ZhangZK2010}
proposed a model to explain some statistical properties in
folksonomy, in which users can search for resources via tags. 
Many of these studies consider individual tag assignment, while
ignoring the interaction among peer tag users.

In this paper, we propose a model to investigate the dynamics and
interaction among individuals in a tagging system. 
Specifically,
individuals imitate each other in tagging which results in a self-organized
state. We found that when users are of low confidence, they
self-organize to attain a steady state of active tagging. On the
other hand, the system ends with inactive tagging when users 
are confident of their own tagging practice. In addition, a phase
transition is observed with a critical level of user confidence,
when the system changes from one regime to the other. 
Furthermore, 
we compare distributions of post length from the proposed
model to two real datasets obtained from \emph{delicious.com} and \emph{flickr.com}, 
which show good agreements.

\section{Model}
\label{sec_model}

We consider a model of tagging system with $N$ users.
At each step,
each user posts one resource and assign tags to the resource.
The tendency of which the user assigns tags 
is characterized by $p_i(t)$,
which is the probability that the user continues with tag assignment for the resource.
In other words,
the probability that user $i$ assigns $n_i(t)$ tags at time $t$ is given by
\begin{eqnarray}
\label{eq_nit}
    \Pr[n_i(t)=l] = p_i^l(t)[1-p_i(t)],
\end{eqnarray}
where $l=1,2,3,\dots$.
Large $p_i(t)$ corresponds to a high tendency to assign tags
and vice versa.
We thus call $p_i(t)$ the {\em tagivity},
which characterizes the tendency of user $i$ in tag assignment.
Given that $p_i(t)$ remains unchanged,
$n_i(t)$ follows a geometric distribution with parameter $1-p_i(t)$.
We model the self-organization of user by assuming that users adjust their $p_i(t)$
based on the observation of $\avg{p(t)}$,
the average tagivity over all users at time $t$.

As one main purpose for tagging is to facilitate the search of resources for others,
users would tend to adopt a more typical tagging practice.
They thus adjust their own tagivity in order to imitate
the observed average tagivity over users.
We denote the combination of tags associated with a resource to be a \emph{post}.
Based on observations,
users obtain an estimated distribution of {\it post length},
which is the number of tags associated with each post.
We assume that the users estimate the distribution based on the average user tagivity,
as given by
\begin{eqnarray}
    \Pr[l'=l] = \avg{p(t)}^l[1-\avg{p(t)}],
\end{eqnarray}
where $l'$ corresponds to the observed post length.
With this distribution in mind,
user $i$ randomly picks a post and imitates its length in the next step.
Suppose user $i$ assigns $n_i(t)$ tags at time $t$,
the probability that he/she picks a post of length $l'$ less than $n_i(t)$
is given by
\begin{eqnarray}
\Pr[l'<n_i(t)]
=1-\avg{p(t)}^{n_i(t)-1}.
\end{eqnarray}
On the other hand,
the probability that user $i$ picks a post of length $l'$ larger than $n_i(t)$
is given by
\begin{eqnarray}
\Pr[l'>n_i(t)]
= \avg{p(t)}^{n_i(t)}.
\end{eqnarray}
With probability $\avg{p(t)}^{n_i(t)-1}(1-\avg{p(t)})$,
user $i$ picks a post of length equals to his/her own post length at time $t$.

Users imitate the post they pick up by changing their tagivities.
For instance,
user $i$ increases his/her tagivities if $n_i(t)$ is smaller than $l'$, 
and vice versa.
We denote the probabilities of which
user $i$ increases, 
maintains or decreases his/her tagivity
as $\eta_i^+(t)$, $\eta_i^0(t)$ and $\eta_i^-(t)$,
given by
\begin{eqnarray}
\label{Eq_p+}
\eta_i^+(t)&=&\frac{(1-\beta)\langle p(t) \rangle^{n_i(t)}}{Z_i(t)},\\
\eta_i^0(t)&=&\frac{\beta[\langle p(t) \rangle^{n_i(t)-1}(1-\langle p(t) \rangle)]}{Z_i(t)},\\
\eta_i^-(t)&=&\frac{(1-\beta)(1-\langle p(t) \rangle^{n_i(t)-1})}{Z_i(t)},
\label{Eq_p+-}
\end{eqnarray}
where $Z_i(t)$ ensures
$\eta_i^+(t)+\eta_i^0(t)+\eta_i^-(t)=1$.
The parameter $\beta\in [0,1]$ can be considered as the {\it confidence} of user on
his own tagivity: $\beta=0$ corresponds to the case with \emph{unconfident}
users who tend to change their choice of tagivities every time step, 
and $\beta=1$ corresponds to the case with \emph{confident} users who
stay with their tagivities every time step. 
Increasing $\beta$ from 0 to 1 characterizes the increase in user confidence,
such that users are more reluctant to changes.

We propose two response functions based on which the tagivity is updated.
In the first case,
the tagivity is updated {\it linearly} by
\begin{eqnarray}
\label{eq_linear}
p_i(t+1) = p_i(t) + a_i(t)\delta_l,
\end{eqnarray}
where $a_i(t)=1$, $0$, $-1$ respectively with probabilities
$\eta_i^+(t)$, $\eta_i^0(t)$, $\eta_i^-(t)$, and $\delta_l>0$ is a
parameter which characterizes the extent the tagivity is
changed. When $a_i(t)=1$ or $-1$, the tagivity increases or decreases. 
The parameter $\delta_l$ can be interpreted as the {\it
adaptability} of the users. Large $\delta_l$ corresponds to faster
adaptation to the typical behaviors.

In the second case, 
the complementary tagivity $1-p_i(t)$ is updated {\it multiplicatively}
by
\begin{eqnarray}
\label{eq_multi}
1-p_i(t+1) = [1-p_i(t)](1+\delta_m)^{-a_i(t)},
\end{eqnarray}
where $\delta_m\ge 0$ serves the same role as $\delta_l$ in linear
update. A more explicit implication of this multiplicative
updating can be obtained by the relation $E[n_i(t)] =
(1-p_i(t))^{-1}$, where $E[n_i(t)]$ is the expected value of
$n_i(t)$ based on the geometric distribution. Equation (\ref{eq_multi}) thus implies
\begin{equation}
E[n_i(t+1)]=E[n_i(t)](1+\delta_m)^{a_i(t)}.
\end{equation}
In other words,
the expected value of $n_i(t)$
respectively increases by a factor of $(1+\delta_m)$,
remains unchanged or decreases
by a factor of $(1+\delta_m)^{-1}$ with $a_i(t)=1, 0 -1$.

\section{Simulation results}

To reveal the dynamics underlying self-organization in the model, 
we conduct numerical simulations. 
We start with random initial $p_i(0)$
for all users. At time $t$, $n_i(t)$ is drawn according to the
probabilities in \req{eq_nit}, such that $\eta_i^+$, $\eta_i^0$ and
$\eta_i^-$ are evaluated according to 
Eqs.~(\ref{Eq_p+})-(\ref{Eq_p+-}). 
The tagivity $p_i(t)$ for each user
is then updated according to \req{eq_linear} in the case of linear
update or \req{eq_multi} in the case of multiplicative update.
Unless specified, the results are obtained when the system
converges, i.e. $\avg{p(t)}$ becomes steady. 
We observed that
$\avg{p(t)}$ has a slight fluctuation around a time average value and the
fluctuation is dependent on $\delta_l$ and $\delta_m$.

 \subsection{Convergence time}

We first study the relation between the convergence time 
and the parameters $\beta$, $\delta_l$ and $\delta_m$. 
The self-organized state in our context corresponds to
the state in which $\avg{p(t)}$ becomes steady. 
The convergence time
$\tau$ is thus defined by the relation $\avg{p(\tau)}\approx\avg{p(\tau+L)}$, 
for all $t\ge\tau$ and some sufficiently large $L$.

\begin{figure}[htb]
 \centering
 \includegraphics[width=6cm]{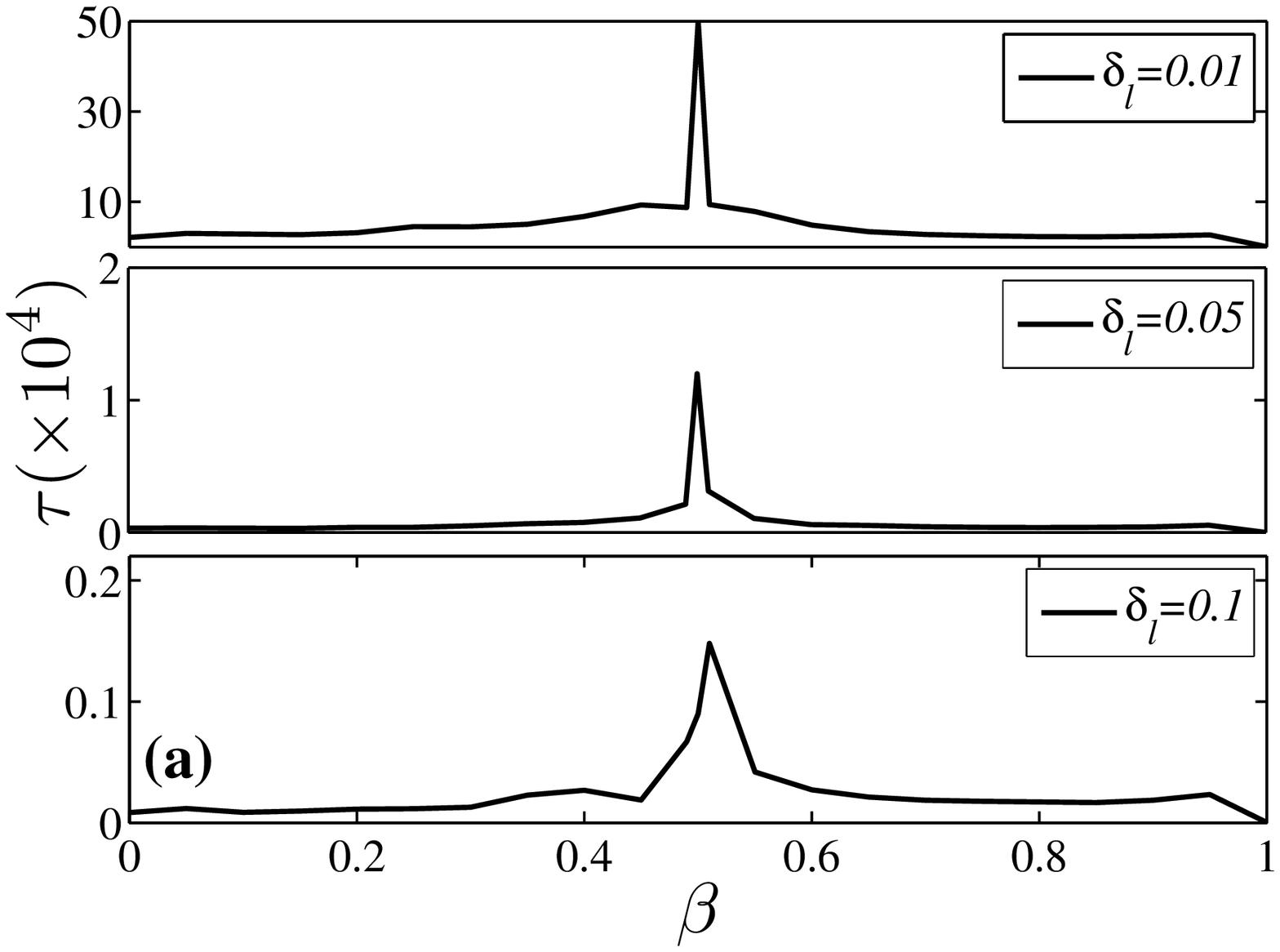}
 \includegraphics[width=6.5cm]{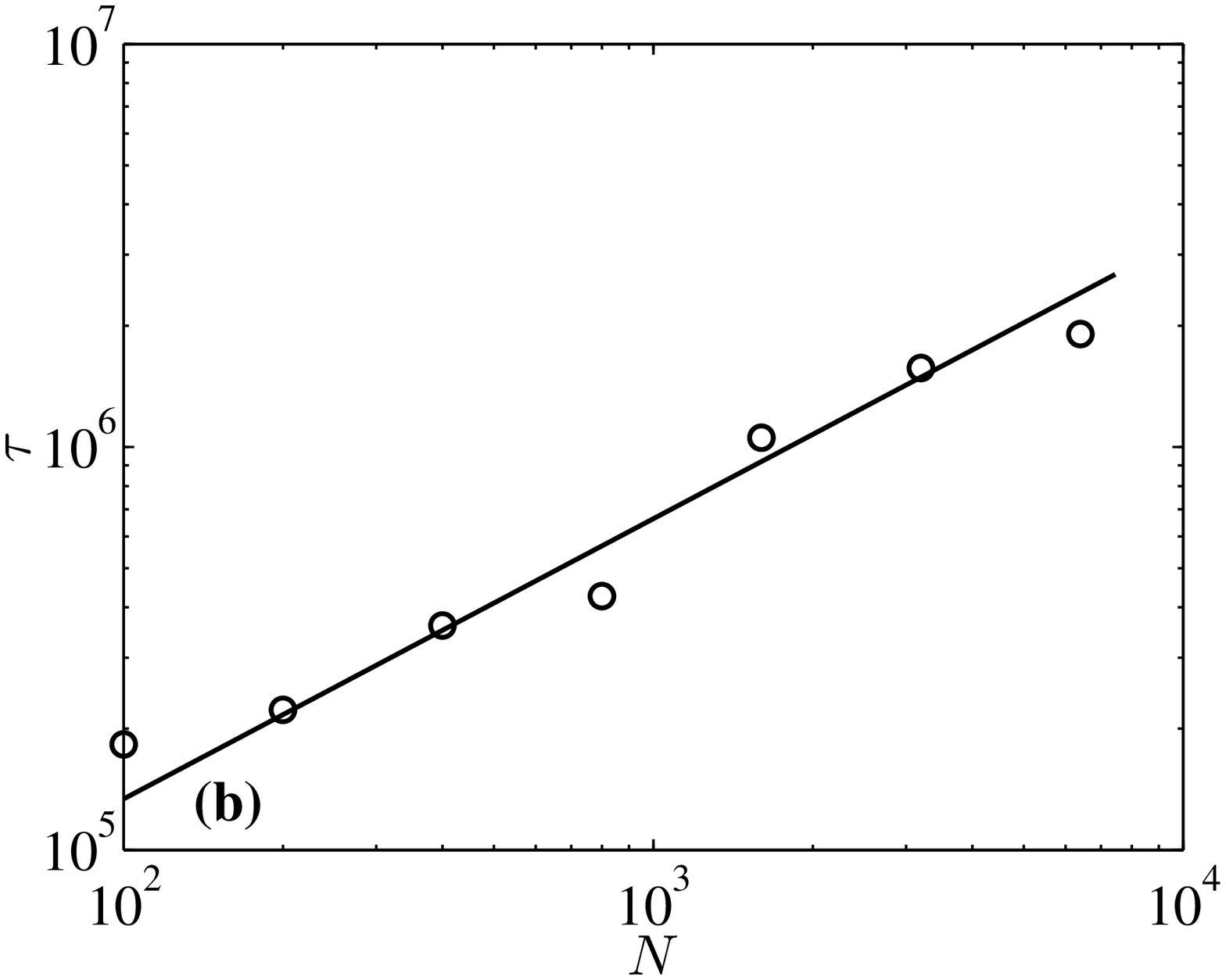}
 \caption{\label{Fig_convergence_time}
 (a) Convergence time $\tau$ as a function of $\beta$ for different $\delta_l$.
Convergence time peaks around $\beta=0.5$.
The larger the adaptability
 $\delta_l$, the more quickly the system reaches steady state.
 (b) Convergence time as a function of $N$ when $\beta=0.5$, which show scattering of data
 round the straight line implying $\ln \tau \propto\ln N$. }
 \end{figure}

The convergence time is
plotted in Fig.~\ref{Fig_convergence_time}(a) as a function of confidence $\beta$.
As similar results are obtained from the two update rules,
we present only the results obtained from the linear update.
As shown in Fig.~\ref{Fig_convergence_time}(a),
the larger the adaptability $\delta_l$,
the faster the convergence time.
The prominent peaks of convergence
time observed at $\beta\approx 0.5$ suggest the possibility of a phase
transition at $\beta\approx 0.5$ as dynamics slows down.
Furthermore,
peak positions are similar at different values of $\delta_l$.
It implies that,
when the weight $\beta$ in 
Eqs.~(\ref{Eq_p+}) - (\ref{Eq_p+-}) to modify tagivity
is equal to that to maintain tagivity, 
the users are confused and the self-organization
slows down. 
As the convergence time is also dependent
on system size,
we plot in log-log scale $N$ as a function of
$\tau$ at $\beta=0.5$ in Fig.~\ref{Fig_convergence_time}(b),
as studies \cite{Holme2006,Medvedyeva2003} suggest a
conventional scaling of $\ln \tau \propto \ln N$ in the proximity of
phase transition.  
These results suggest that on top of the self-organization, there is a phase
transition in the range close to $\beta=0.5$.

 \subsection{Steady distributions of tagivity}

\begin{figure}[htb]
 \centering
 \includegraphics[width=6cm]{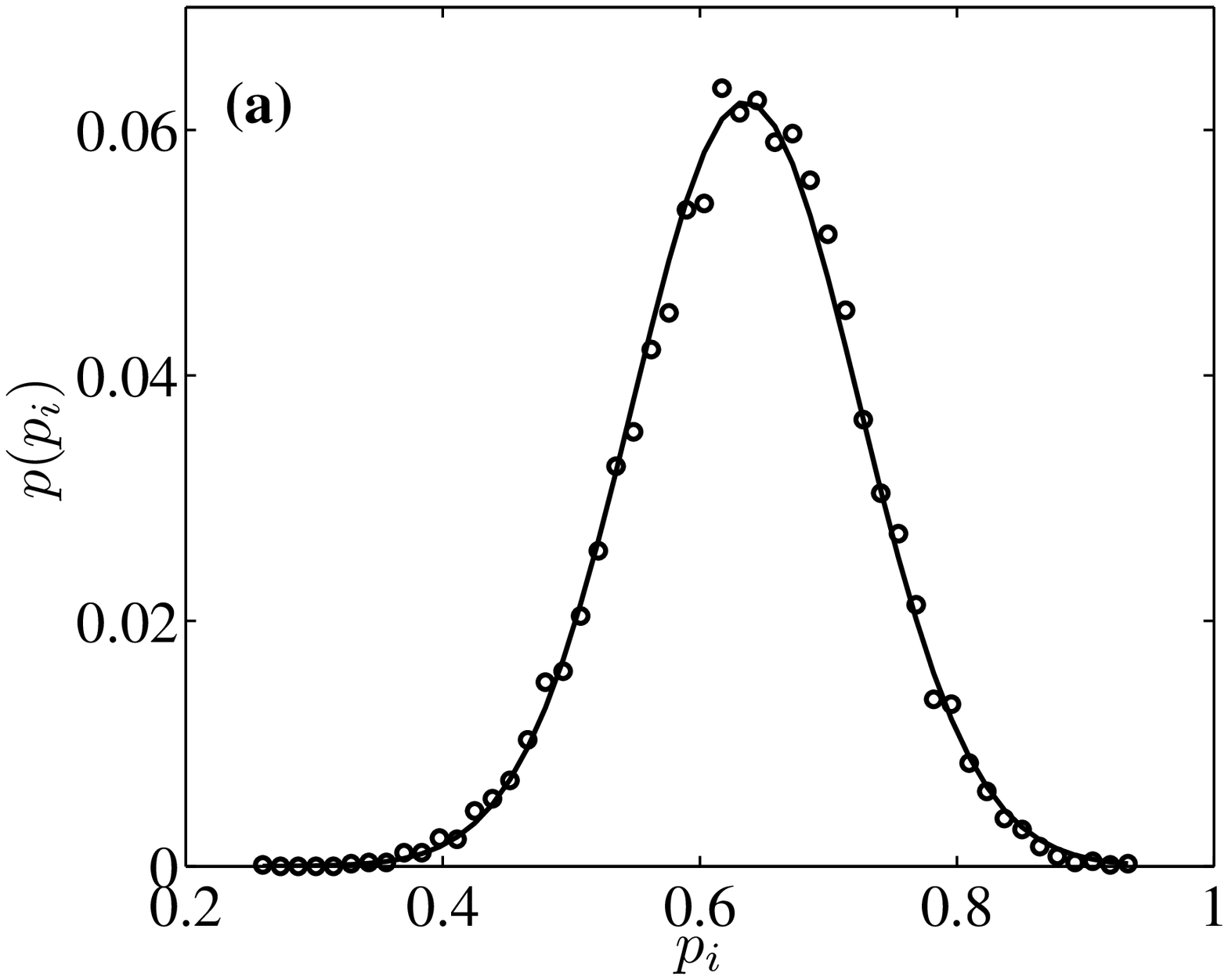}
 \includegraphics[width=6cm]{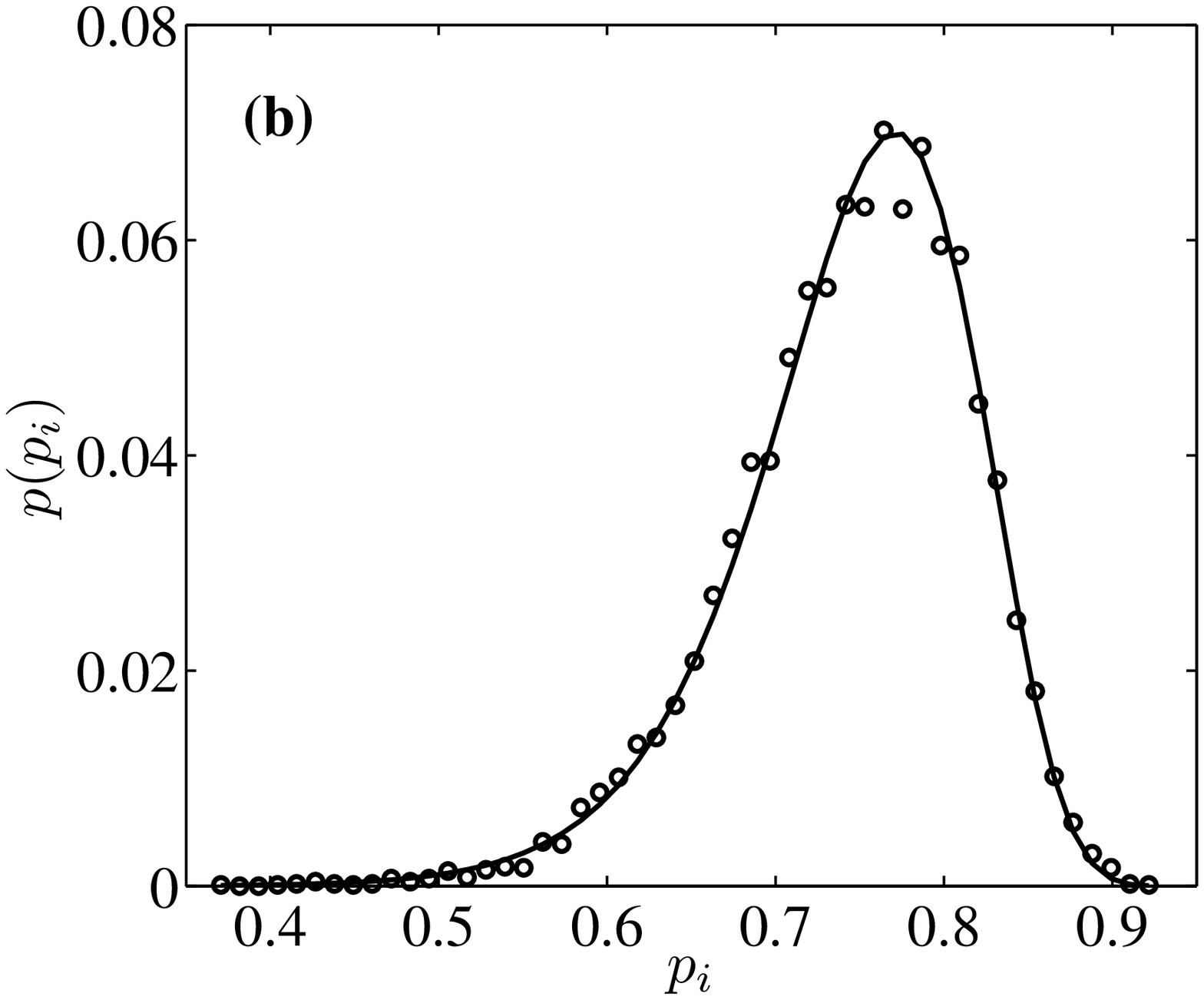}
 \caption{\label{Fig_distribution_p} The tagivity distributions with
 (a) the linear update and (b) the multiplicative update.
 Parameters: $\beta=0.45$ and $\delta_l=0.05$ for linear update
 and $\beta=0.4$ and $\delta_m=0.1$ for multiplicative update.
 Fittings: (a) Gaussian fit with  $\mu=0.63$
 and $\sigma=0.088$ and (b) log-normal fitting with $\mu=-1.41$ and $\sigma=0.27$. }
 \end{figure}

As mentioned in Sec.~\ref{sec_model},
each user at each step randomly picks a post and imitates its length,
their tagivities thus fluctuate around the average values.
We show in \fig{Fig_distribution_p} the stable distribution of tagivity
after the system converges.
Figure~\ref{Fig_distribution_p}(a) shows that the stable distribution from linear update
resembles Gaussian distribution.
The simulation results are obtained by $\beta=0.45$ and $\delta_l=0.05$
and the parameter of Gaussian fit are $\mu=0.63$ and $\sigma=0.088$.
The results are not as obvious as \req{eq_linear} suggests in the case when $a_i(t)$
is a random variable.
In such case,
$\sum_{t=1}^{\infty}a_i(t)$ would result in an infinite variance of $p_i(t)$,
as compared to the finite variance observed in \fig{Fig_distribution_p}(a).
The finite variance of $p_i(t)$ comes from the restoring process of $a_i(t)$
around the typical behaviors,
as given by the probabilities in 
Eqs.~(\ref{Eq_p+})~-~(\ref{Eq_p+-}). 
Figure~\ref{Fig_distribution_p}(b)
shows the stable distribution of tagivity obtained from the
multiplicative update, 
where $1-p_i$ approximately follows the
log-normal distribution. Simulation results are obtained by
$\beta=0.4$ and $\delta_m=0.1$, with log-normal fitting of
$\mu=-1.41$ and $\sigma=0.27$. The origin of the log-normal
distribution is similar to that of Gaussian distribution and can be
seen by taking algorithm of \req{eq_multi}.

\section{Phase transition and Self-organization}

Though analytic solutions for the general case are difficult to obtain,
we can write down a simple description of the steady state when $\delta_l\rightarrow 0$ or $\delta_m\rightarrow 0$.
In this case
we assume $p_i\approx\avg{p}$ for all user $i$.
We further introduce a quantity $\Delta$
which characterizes the  tendency for $\avg{p}$ to increase or decrease,
as given by
\begin{eqnarray}
\label{eq_delta}
    \Delta(\avg{p}) = \sum_{n=1}^{\infty}\avg{p}^{n-1}(1-\avg{p})[\eta^+(n,\beta)-\eta^+(n,\beta)],
\end{eqnarray}
$\Delta$ describes the difference between $\eta^+$ and $\eta^-$
when the average user tagivity is $\avg{p}$.
A positive $\Delta$ corresponds to a tendency for $\avg{p}$ to increase,
and vice versa.
Substitutions of 
Eqs.~(\ref{Eq_p+}) and (\ref{Eq_p+-}) for $\eta^+$ and $\eta^-$ into \req{eq_delta} lead
to the following expression
\begin{equation}
\label{eq_delta2}
  \Delta(\avg{p})=\sum_{n=1}^{\infty}\frac{\langle p \rangle^{n-1}(1-\langle p \rangle)(\langle p \rangle^{n}-1+\langle p \rangle^{n-1})}{Z(n, \beta)}.
\end{equation}
We numerically evaluate the summation in \req{eq_delta2} 
and obtain the values of $\avg{p}$ when $\Delta=0$,
i.e. when there is no tendency for $\avg{p}$ to increase or decrease
and the system becomes steady.

 \begin{figure}[htb]
 \centering
 \includegraphics[width=7cm]{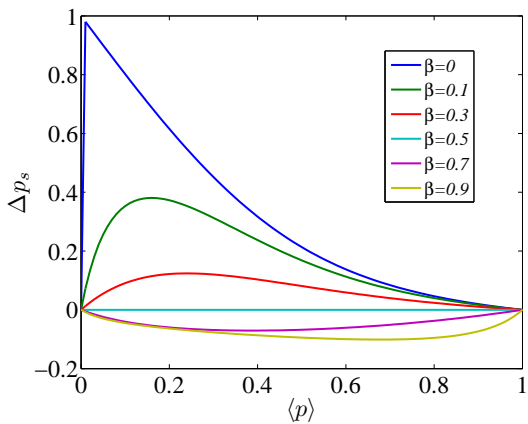}
 \caption{\label{Fig_p_dp_beta} (Color online) $\Delta$ as a function of $\avg{p}$ for different values of $\beta$.}
 \end{figure}

Figure~\ref{Fig_p_dp_beta} shows $\Delta$ as a function of $\avg{p}$ for different  $\beta$.
These results imply that for all $\beta$,
$\avg{p}=0,1$ are solutions of $\Delta=0$.
The fixed points of $\avg{p}=0$ or $\avg{p}=1$
respectively correspond to the cases when all users in the system stop active tagging
or assign infinite number of tags.
When $\beta<0.5$, we get $\Delta\ge 0$ for all $\avg{p}$,
which implies that the tendency to increase tagivity is higher than that to decrease tagivity,
leading to a stable fixed point at $\avg{p}=1$.
On the other hand, 
when $\beta>0.5$, 
we get $\Delta\le 0$ which implies that
the tendency for the tagivity to decrease is larger than that to increase,
leading to an opposite result of stable fixed point at $\avg{p}=0$.
This drastic change of the self-organized state corresponds to a phase transition at $\beta\approx 0.5$
from a regime with active tag assignment to one with inactive tag assignment.
It is also interesting to note that when $\beta=0.5$,
$Z\equiv 1$ for all $n$ in \req{eq_delta2} such that $\Delta \equiv 0$ is guaranteed by the identity
  \begin{equation}
  \sum_{n=1}^{\infty} \langle p \rangle^{n-1}(1-\langle p \rangle^{n-1}-\langle p \rangle^n)\equiv 0
  \label{Eq_p_simplify}
  \end{equation}
for all values of $\avg{p}$. It implies
that at the critical point of $\beta=0.5$, the system does not have
a unique fixed point of $\avg{p}$, unlike the cases with
$\beta\neq 0.5$.

 \begin{figure}[htb]
 \centering
 \includegraphics[width=7cm]{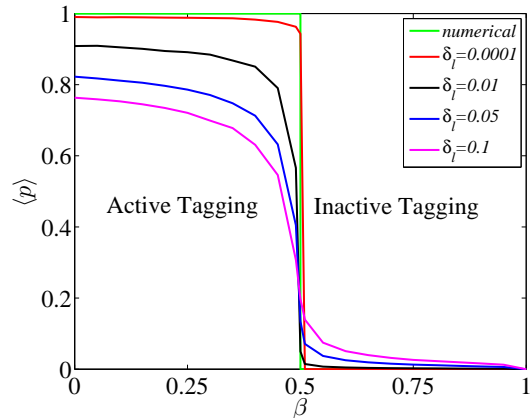}
 \caption{\label{Fig_averagep_beta} (Color online) The average tagivity $\langle p \rangle$ as a function of
 $\beta$ for various $\delta_l$. The analytical results at $\delta_l \rightarrow 0$ is shown by the green line.}
 \end{figure}

These analytical predictions of $\avg{p}$ with $\delta_l=0$ are compared to
simulation results with $\delta_l>0$. 
As the results obtained from the two update rules are similar, 
we present only the results obtained
from the linear update. 
The \emph{green line} in
\fig{Fig_averagep_beta} shows the analytical stable fixed points of $\avg{p}$ with $\delta=0$. 
We find that simulations with small $\delta_l$ agrees well with the
analytical limit, and for $\delta_l>0$, $\avg{p}$ decreases with
increasing $\beta$ as well as increasing $\delta_l$. 
As we can see,
all the simulation results show an abrupt change in $\avg{p}$ at
$\beta\approx 0.5$, 
suggesting the existence of a phase transition
as predicted by the analytical results.
We remark that $\beta=0.5$ corresponds to the case in 
Eqs.~(\ref{Eq_p+})~-~(\ref{Eq_p+-}) where the weight to
imitate others equal to that to stay unchanged.
These results imply that when users have low
confidence, they tend to imitate each other in tagging which leads to
a steady state of active tag assignment. However, when users are confident
and are stubborn for changes, they stay with their own practice and
result in a steady state with inactive tagging.  
These two behaviors are connected by an abrupt change when confidence
increases across $\beta=0.5$.

 \begin{figure}[htb]
 \centering
 \includegraphics[width=7cm]{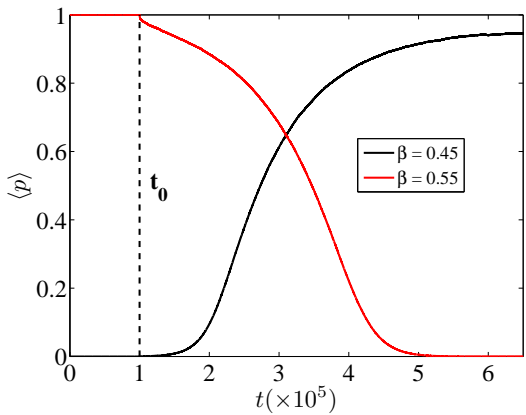}
 \caption{\label{Fig_p_time} (Color online) The dynamics of which the self-organization is established.
 \emph{Black line}: $p_i(0)=0$ for all users, and at time $t_0$ one user
 assigns more than one tags.
 \emph{Red line}: $p_i(0)=1$ for all users, and at time $t_0$ one user
 assigns only one tag.}
 \end{figure}

To show explicitly how users self-organize to attain the steady state, we
start the system at the unstable fixed point and examine how it evolves to the stable fixed point
after a slight perturbation.
The black line in \fig{Fig_p_time} corresponds to the average tagivity
for the case when confident users (i.e. $\beta<0.5$) are initialized with zero tagivity. 
At time $t_0$, one of the users
assigns a tag which initiates others to imitate. 
As we can see, 
the average tagivity slowly increases after $t_0$ and saturates at a non-zero
steady value, correspond to the self-organization from inactive
to active tagging.
On the contrary, the red line shows the case when users are initialized
with $p_i(0)=1$ and large confidence (i.e. $\beta>0.5$). A maximum
post length is set to avoid infinite tagging. At time $t_0$, 
one user assigns the minimum number of tags which initiates others
to imitate. As we can see, the average tagivity slowly decreases
after $t_0$ and becomes steady at zero, 
corresponds to the self-organization from active to inactive tagging.

\section{Empirical results}

As it is difficult to define and obtain the tagivity for real
users, 
other well-defined quantities are used for comparison.
We compare the distributions of post length obtained from the model with two
real datasets: (1)
\emph{delicious.com}, a social bookmarking website for saving,
sharing and discovering bookmarks associated with tags;
(2) \emph{flickr.com}, an image hosting website which encourages
users to organize their pictures with tags.

 \begin{figure}[htb]
 \centering
 \includegraphics[width=7cm]{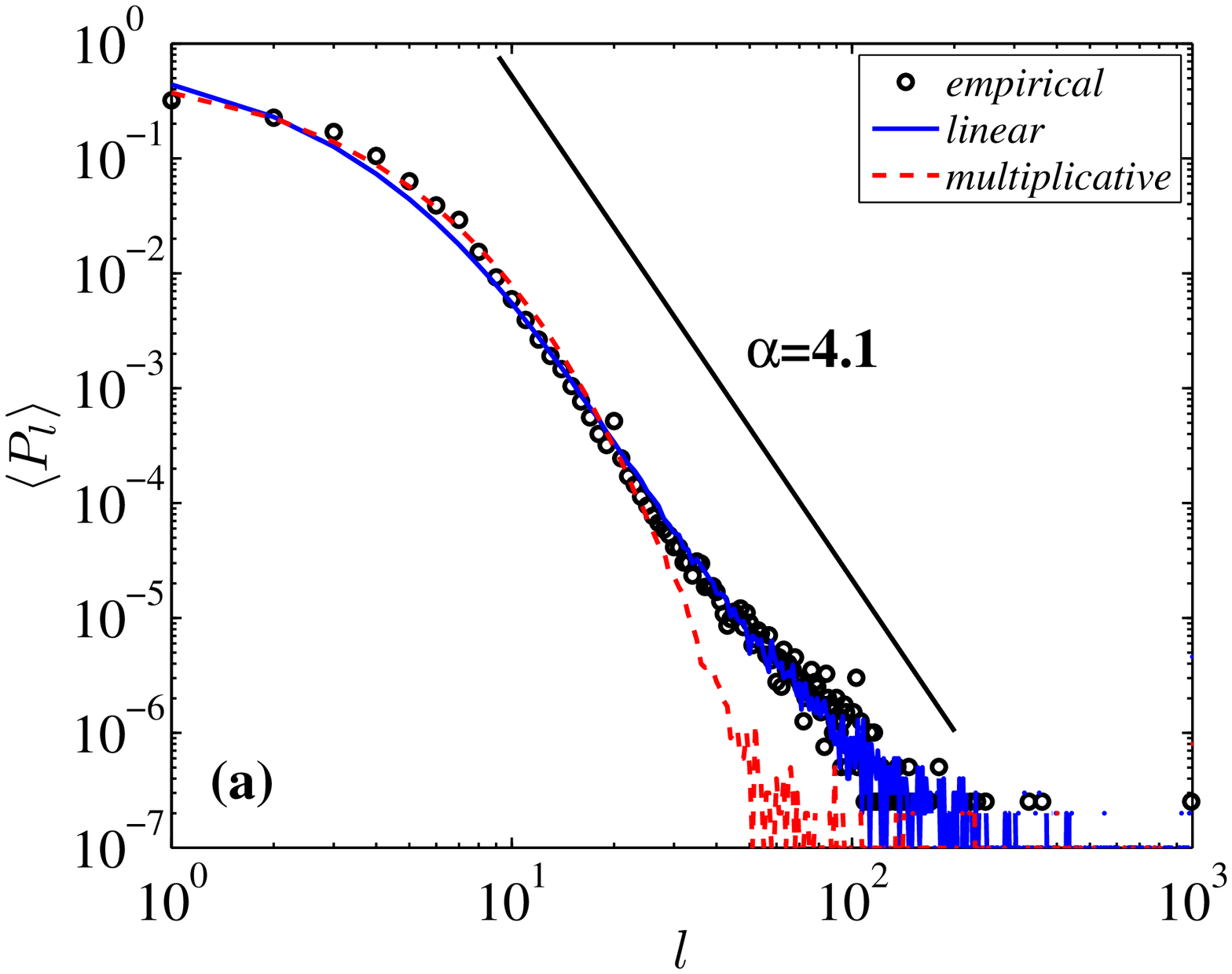}
 \includegraphics[width=7.5cm]{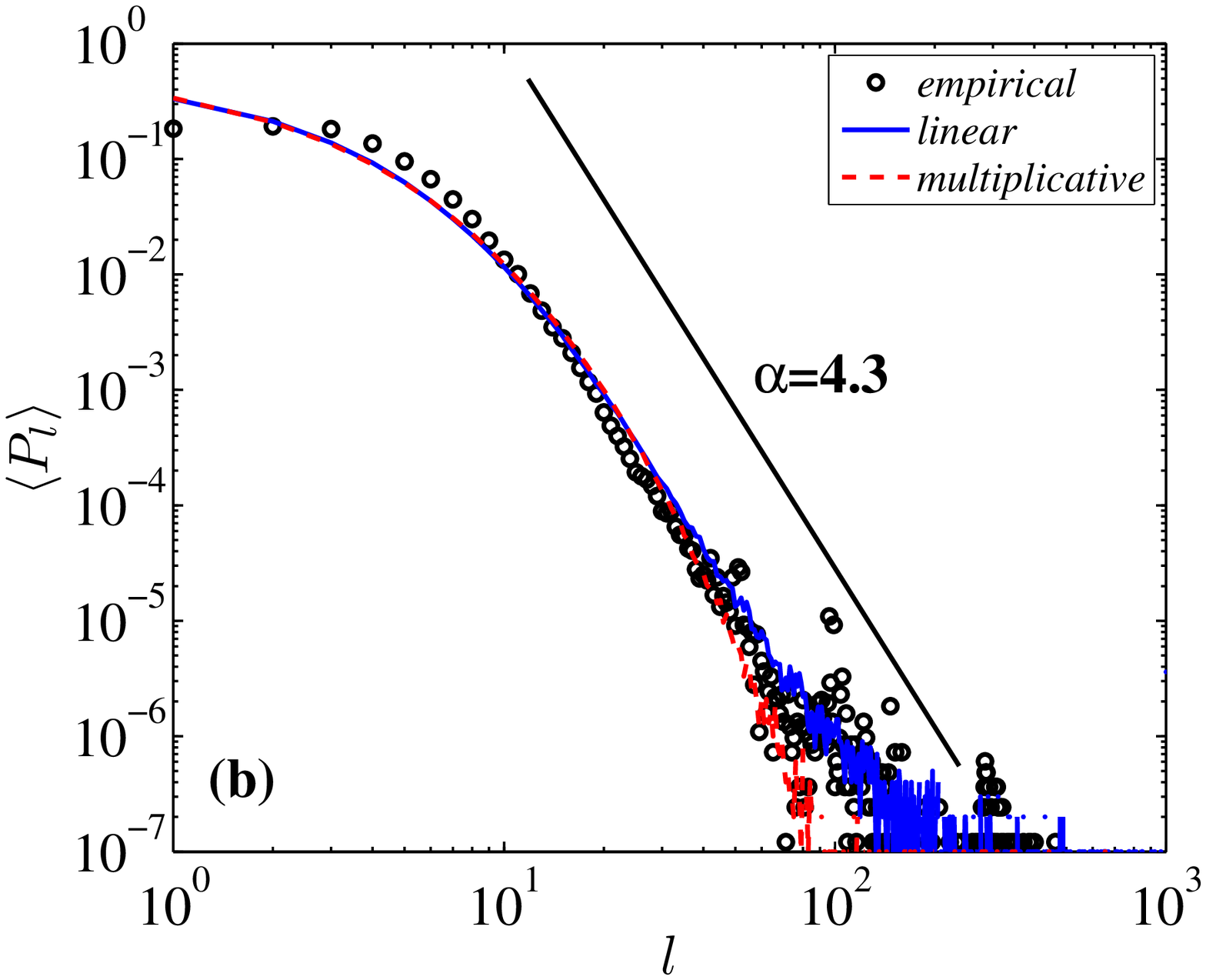}
 \caption{\label{Fig_empirical_distribution} (Color online) Empirical distribution
 of the number of tags in each post as compared to simulations. \emph{Circles} represent empirical
 data, \emph{blue solid lines} and \emph{red dash lines} respectively represent simulation results with linear
 and multiplicative update. (a) Data from \emph{delicious.com} compared to simulations with parameters 
$\beta=0.45$ and $\delta_l=0.08$ for linear update and
$\beta=0.45$ and $\delta_m=0.1$ for multiplicative update.
(b) Data from \emph{flickr.com} compared to simulations with
parameters $\beta=0.4$, $\delta_l=0.06$ for linear update and $\beta=0.4$, $\delta_m=0.2$ for multiplicative update.}
\end{figure}

We show in Fig.~\ref{Fig_empirical_distribution} (a) and (b) the distributions
of the post length (as open circles) obtained respectively from \emph{delicious.com} and \emph{flickr.com}.
The posts without tags are removed from the statistics.
 It is interesting to note that the two distributions display similar behaviors:
 an initial
 fast decay with post length less than $8$, followed by a power-law decay for intermediate post length,
 and then a high tail.
The exponents of the power law decay are $4.1$ and $4.3$ respectively in \emph{delicious.com} and \emph{flickr.com},
with average post length approximately 2.9 and 3.4.
The simulated distributions
are plotted in Fig.~\ref{Fig_empirical_distribution}
as \emph{blue} and
\emph{red} lines respectively for linear and multiplicative update,
all with $\beta<0.5$.
These results may suggest that real users are of low confidence and tend
to imitate each other in tag assignment.

As we can see,
the simulation results based on the linear update have better agreement
with empirical data than that of the multiplicative update.
With the linear update,
the high tails of empirical data are also well fitted.
According to \fig{Fig_distribution_p},
the tagivity distribution obtained from the linear update shows a slower decay at large $p$,
as compared to the faster decay in the multiplicative case.
The slow decay at large $p$,
i.e. more users are found with large tagging tendency,
may explain the high tail in the post length distributions.

\section{Conclusions and Discussion}

In this paper, we proposed a model to illustrate the
self-organization of tagging behaviors in social tagging systems,
where individuals imitate each other in tag assignment and eventually result in a self-organized state.
With linear update on the tagging tendency, namely \emph{tagivity},
the corresponding steady distribution resembles Gaussian
distribution. On the other hand, the steady distribution resembles
log-normal distribution when multiplicative update is employed.
In addition, we found that when users are of low confidence,
they tend to imitate others and the system ends with a steady state of active tagging.
By contrast, when users are of high confidence, the system will reach a steady state of inactive tagging.
Abrupt changes are observed when user confidence increases and the
system changes from one regime to the other, suggesting a phase
transition separating the active and inactive tagging.
Analyses on convergence time suggest a slow dynamics around the parameter
range of phase changes, 
which provides further evidence for the
transition. Finally, the post length distributions of the model are compared to two
real datasets obtained from \emph{delicious.com} and \emph{flickr.com},
which show good agreements.

Social tagging systems have been studied with approaches 
ranging from graph theory to statistics,
which may overlook the interactions and dynamics among individuals.
The present model introduced in this paper provides
a simple yet interesting description of evolving social tagging systems,
which might be generalized to other systems where self-organizations are observed.
The proposed model may also shed light on applications (e.g. recommender systems \cite{ZhangZK20102, ZhangZK20103})
which combine statistical physics and agent-based models \cite{Bonabeau2002}
in understanding tagging systems as well as other social systems \cite{Castellano2000}.

\section*{Acknowledgment}

This work was partially supported by the Program for New Century
Excellent Talents in University (NCET-07-0288), the Fundamental
Research Funds for the Central Universities
and QLectives projects (EU FET-Open
Grants 213360 and 231200). ZKZ acknowledges the National Natural Science Foundation of China (Grant nos. 60973069 and 90924011).

\end{document}